# The First Order Signal in Pure $U(1)$ Gauge Theory May be Fake [*]

C. B. Lang[a] and T. Neuhaus[b]

[a]Institut für Theoretische Physik, Universität Graz, A-8010 Graz, Austria

[b]Fakultät für Physik, Universität Bielefeld, 33615 Bielefeld, Germany

We study the deconfinement phase transition of compact $U(1)$ pure lattice gauge theory with the Wilson action on *closed topology* lattices. In contrast to studies of compact $QED$ on *hypercubic lattices with periodic boundary conditions*, we find no metastability signal at the phase transition on the lattices with the topology of a sphere. Thus the determination of the order of this phase transition has to be reconsidered. We argue that different properties of closed monopole loops on these topological inequivalent lattices might be responsible for the effect.

## 1. INTRODUCTION

For some time now computer studies of the phase transition of compact $U(1)$ pure gauge theory with the Wilson action have produced first order signals. Original work [1] on comparatively small lattices could not resolve the energy gap, but subsequent high statistics Monte Carlo results exhibited a metastability signal [2,3]. Detailed studies of the latent heat in an extended parameter space corresponding to the adjoint representation of the plaquette pointed towards the existence of a tricritical point at negative values of the adjoint coupling [4]. It was then generally accepted that the compact $U(1)$ pure gauge theory with the Wilson action has a weak first order PT with a small latent heat $\Delta\langle\cos(U_P)\rangle \approx 0.016$ in the thermodynamic limit.

It was observed quite early [5] that this PT is accompanied by an intriguing behaviour of topological excitations, so-called monopoles. These are artifacts of the compact formulation of the $U(1)$-action, i.e. flux deficits in 3-cubes. On the dual lattice they form closed loops due to current conservation. In fact, suppressing them affects the PT, moving it to the disordered confinement phase [6]. Moreover eliminating monopoles from the path integral by introducing certain constraints makes the confinement phase disappear for positive values of $\beta$ [7].

Almost all numerical simulations have used periodic boundary conditions on hypercubic lattices, some have chosen helical b.c., which do not behave differently with regard to the subject discussed here. Hypercubic lattices with periodic boundary conditions are homotopic the surface of a torus with the fundamental homotopy group $\pi_1 \simeq \mathbf{Z} \oplus \mathbf{Z} \oplus \mathbf{Z} \oplus \mathbf{Z}$. The formulation of compact $QED$ on the torus contains certain peculiarities with respect to the monopole excitations of the theory: A monopole loop (on the dual lattice) may be considered as the boundary of a Dirac sheet. Whereas the shape of the Dirac sheet is gauge variant its boundary is not. A pair of monopole loops wrapping around the torus may be connected by a Dirac sheet and pairwise annihilation may leave a closed Dirac sheet in the lattice configuration. This has been observed indeed on finite lattices as a metastable energy gap in the Coulomb phase of the model [8]. Furthermore due to the periodic b.c. one finds monopole loops that are closed only because of the periodic boundary conditions and that may not be continuously deformed (are not homotopic to) to points. They are present in the "hot" phase of the model, but usually not in the "cold" phase. These particular properties of the theory on a torus might have led to a wrong identification of the order of the $QED$ deconfinement PT with the Wilson action. In fact earlier MCRG studies of the model showed in the vicinity of the Wilson line a flow of couplings, which resembled the flow diagram of a second order critical point quite well [9].

[*]Work supported by DAAD and ÖAD

## 2. LATTICE WITH TRIVIAL HOMOTOPY GROUP

The central idea is to study compact QED with the Wilson action on a 4D lattice, which has the topology of a sphere $\pi_1(S_4) \simeq \mathbf{1}$. On such a lattice any closed monopole loop is homotopic to a point and therefore above mentioned complications due to the periodic boundary conditions are excluded. For the construction we choose the $D$-manifold as the boundary of a $D+1$-manifold, for which we select a $D + 1$-dimensional hypercube $H_{D+1}(L)$ with linear extend $L$ and $L^{D+1}$ sites. We denote its boundary by $SH_D(L)$; it consists of all sites which have at least one cartesian coordinate value 1 or $L$. In Table 1 we display the number of basic elements (sites, links, plaquettes) of the manifold $SH_4(L)$. Ratios of e.g. $n_{\text{links}}/n_{\text{sites}}$ approach for the $SH_4(L)$ lattice for large $L$ asymptotically the values for the hypercubic geometry $H_4(L)$. Whereas $H_D(L)$ is self-dual, $SH_D(L)$ is obviously not. It is however possible to construct the dual manifold $SH'_D(L)$ (which has the same homotopy properties) and we have used this construction for some exploratory identifications of closed monopole loops.

Table 1
Characteristic numbers on $SH_4(L)$ lattices.

| manifold | $SH_4(L)$ |
|---|---|
| sites | $10(L-1)^4 + 20(L-1)^2 + 2$ |
| links | $40(L-1)^4 + 40(L-1)^2$ |
| plaquettes | $60(L-1)^4 + 20(L-1)^2$ |

## 3. SIMULATION AND RESULTS

We have simulated compact QED with the Wilson action on lattices $SH(4), SH(6), SH(8)$ and $SH(10)$. Comparing the number of degrees of freedom, i.e. the number of link variables, we find that those systems roughly correspond to hypercubic lattices $H(6), H(9), H(12)$ and $H(16)$ in $D = 4$. The geometrical properties have been implemented by tables and we use a 3-hit Metropolis update.

We have studied long runs of $O(10^5)$ sweeps at various couplings $\beta$ close to the deconfinement PT. We then combine our data using the Ferrenberg-Swendsen technique and obtain an optimal estimator for the density of states. From that we determine various quantities like the internal energy, the specific heat $C_V(\beta, L)$ and the BCL-energy-cumulant $V(\beta, L)$.

Naively one expects any differences due to the topology of the lattice to disappear in the large volume limit. Therefore we expect similar pseudocritical coupling estimators for our case than in the case of the torus topology. The pseudocritical couplings as defined by the position of the maxima of the specific heat as a function of $L$ are $\beta_c(4) \approx 1.0155$, $\beta_c(6) \approx 1.0135$, $\beta_c(8) \approx 1.0127$ and $\beta_c(10) \approx 1.0122$. Simulations for the theory on the hypercubic lattice with periodic boundary conditions show clear two-state signals with a pronounced double peak distribution in the energy near the PT. States with energies in between the pure phases are clearly suppressed. In our runs, however, we could not find any such signals. Away from the pseudocritical points the energy distribution histograms always were purely Gaussian, while at pseudocriticality their shape somewhat departs form a Gaussian form. Fig. 1 compares the energy distribution functions of a run on the $SH(8)$ lattice with a run on a $H(10)$ lattice. In both runs the couplings were chosen very close to the peak locations of the specific heat.

We did however observe very long autocorrelation times up to $O(1000)$ sweeps for our largest lattices at pseudocriticality. We also analyzed the finite size scaling properties of the pseudocritical couplings $\beta_c(L)$ as well as the scaling properties of the maximum of the specific heat $C_V^{max}(L)$ and the minimum of the BCL-energy-cumulant $V^{min}(L)$. Fitting these quantities in the $L$-interval $6 \leq L \leq 10$ with the large $L$ scaling laws

$$\begin{aligned}
C_V^{max}(L) &= aL^{\frac{\alpha}{\nu}} \\
V^{min}(L) &= V^{min} - bL^{-4} \\
\beta_c(L) &= \beta_c + cL^{-\frac{1}{\nu}}
\end{aligned}$$

we obtain the following rough estimates for $\frac{\alpha}{\nu} \approx 3.1$, $\nu \approx 1.0$, $V^{min} \approx 0.66655 < \frac{2}{3}$ and $\beta_c \approx$ 1.0102. These estimates should be taken with precaution. It might well be that the asymptotic scaling behaviour for above finite size scaling laws sets in only on very large lattice sizes. While in general not much is known about this for gauge theories, simpler models, like e.g. $D = 2$ Potts models require lattices as large as 10 times the finite correlation length at the first order PT in order to show the above asymptotic scaling behaviour [10]. It is therefore no surprise that those estimates are contradictory among themselves: a value of $V^{min} \neq \frac{2}{3}$ indicates a first order PT, while the value of the $C_V$-exponent $\frac{\alpha}{\nu} \neq 4$ does not.

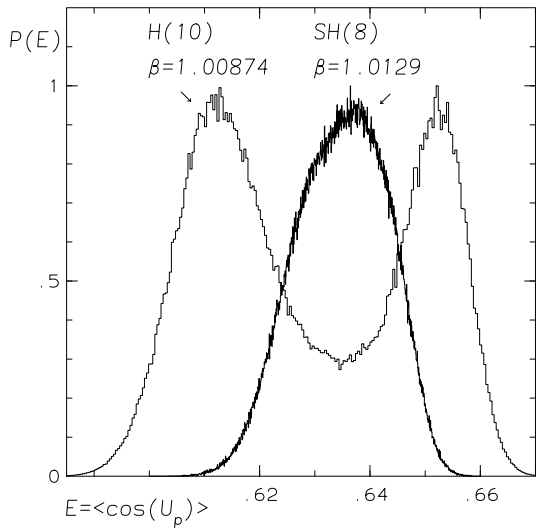

Figure 1. A comparison of energy distribution histograms for compact QED with the Wilson action for the $H(10)$ hypercubic lattice with periodic boundary conditions and the $SH(8)$ closed topology lattice. The simulation on $H(10)$ have been done in the multicanonical Ensemble. The accumulated statistics on the $SH(8)$ lattice is larger than half a million sweeps.

Right now our conclusion has to be that we have not yet sufficient statistics and large enough lattice sizes to unambiguously decide on the order of the PT from the simulations for the Wilson action. However the blunt absence of metastability signal in our data, as compared to the standard case, makes it hard to accept that a first order signal will appear in the thermodynamic limit. After all, first order phase transitions are defined by coexistence of metastable bulk states forming an interface in between them. Our data show no sign of such an interface (cf. Fig. 1). As obviously now there is reason to doubt the usual identification as first order PT we think, that due to the interest on other phase transitions of theories with the compact $U(1)$-gauge group (like QED with matter fields) this problem should be clarified. Like emphasized in the studies which predicted the tricritical point of QED [4], a possible approach to the resolution of the problem would be the inclusion of adjoint plaquette terms into the action. Simulations in this direction are on the way.